\documentclass[12pt,a4paper]{article}
\usepackage{amsmath}
\usepackage{bm}
\usepackage{amsfonts}
\usepackage{graphicx}
\usepackage[normal]{caption2}
\usepackage{subfigure}
\usepackage{rotating}
\usepackage{citesort}
\usepackage{lscape}
\usepackage{section}
\begin{document}
\renewcommand\arraystretch{1.1}
\setlength{\abovecaptionskip}{0.1cm}
\setlength{\belowcaptionskip}{0.5cm}
%%%%%%%%%%%%%%%%%
\title {System size dependence of intermediate mass fragments in heavy-ion collisions }
%%%%%%%%%%%%%%%%%%%%%%%%%%%%%%%%
\author {Sukhjit Kaur \\
 \it House No. 465, Sector-1B, Nasrali, Mandi Gobindgarh-147301, Punjab, India.\\
} \maketitle
%PACS number: 25.70.-z, 25.70.Jj \\
 Electronic address:~sukhjitk85@gmail.com
\begin{abstract}
We simulate the central reactions of $^{20}$Ne+$^{20}$Ne,
$^{40}$Ar+$^{45}$Sc, $^{58}$Ni+$^{58}$Ni, $^{86}$Kr+$^{93}$Nb,
$^{129}$Xe+$^{118}$Sn, $^{86}$Kr+$^{197}$Au and
$^{197}$Au+$^{197}$Au at different incident energies for different
equations of state (EOS), binary cross sections and different
widths of Gaussians. A rise and fall behaviour of the multiplicity
of intermediate mass fragments (IMFs) is observed. The system size
dependence of peak center-of-mass energy E$_{c.m.} ^{max}$ and
peak IMF multiplicity $\langle$N$_{IMF}\rangle^{max}$ is also
studied, where it is observed that E$_{c.m.}^{max}$ follows a
linear behaviour and $\langle$N$_{IMF}\rangle^{max}$ shows a power
law dependence. A comparison between two clusterization methods,
the minimum spanning tree and the minimum spanning tree method
with binding energy check (MSTB) is also made. We find that MSTB
method reduces the $\langle$N$_{IMF}\rangle^{max}$ especially in
heavy systems. The power law dependence is also observed for
fragments of different sizes at E$_{c.m.} ^{max}$ and power law
parameter $\tau$ is found to be close to unity in all cases except
A$^{max}$.
\end{abstract}
\section{Introduction}
 \par
  The behaviour of hot and dense nuclear
matter at the extreme conditions of temperature and density is a
question of keen interest. It can be studied with the help of
heavy-ion reactions at intermediate energies. At high excitation
energies, the colliding nuclei may break into several small and
intermediate size fragments followed by a large number of nucleons
\cite{rkpuri,sood09,jsingh00}. A large number of experimental
attempts had been carried out ranging from the evaporation of
particles to the total disassembly of the dense matter. Besides
these two extremes, one can also have a situation where excited
matter breaks into several fragments. In the last few decades,
several experimental groups have carried out a complete study of
fragment formation with 4$\pi$ detectors
\cite{blaich93,tsang93,sisan01,peas94,desouza91,stone97,ogil91}.
It is quite obvious from these studies that the fragments formed
in heavy-ion collisions depend crucially on the bombarding energy
and impact parameter of the reaction
\cite{rkpuri,sood09,jsingh00,blaich93,tsang93}. Therefore, these
experimental studies of fragmentation offer a unique opportunity
to explore the mechanism behind the formation of the fragments.
Moreover, one can also pin down the role of dynamics in fragment
formation and their time scale.
\par
Recently, there has been
increasing interest in the effects of reaction dynamics on the
production of IMFs and light charged particles (LCPs, Z=1 or 2).
Sisan $\textit{et al.}$ \cite{sisan01} studied the emission of
IMFs from central collisions of nearly symmetric systems using
4$\pi$-Array set up where they found that the multiplicity of IMFs
shows a rise and fall with increase in the beam energy. They
observed that E$_{c.m.}^{max}$ (energy at which the maximum
production of IMFs occurs) increases linearly with the system mass
whereas a power law ($\propto$ A$^{\tau}$) dependence was reported
for peak multiplicity of IMFs with power factor $\tau$ = 0.7.
Peaslee $\textit{et al.}$ \cite{peas94}, on the other hand,
studied asymmetric system $^{84}$Kr+$^{197}$Au in the incident
energy range from 35 to 400 MeV/nucleon and obtained an energy
dependence of multifragmentation. Their findings revealed that
fragment production increases up to 100 MeV/nucleon and then
decreases with increase in incident energy. De Souza $\textit{et
al.}$ \cite{desouza91} studied the central collisions of
$^{36}$Ar+$^{197}$Au from 35 to 120 MeV/nucleon and observed that
IMF multiplicity shows a steady increase with increase in the
incident energy. The IMF multiplicity decreases, however, when one
moves from central to peripheral collisions. On the other hand,
Tsang $\textit{et al.}$ \cite{tsang93}, in their investigation of
$^{197}$Au+$^{197}$Au collisions at E/A = 100, 250, and 400 MeV,
found the occurrence of peak multiplicity at lower energies for
central collisions whereas it is shifted to higher energies for
peripheral collisions. Stone $\textit{et al.}$ \cite{stone97} used
a nearly symmetric system of $^{86}$Kr+$^{93}$Nb from 35 to 95
MeV/nucleon to obtain IMF multiplicity distribution as a function
of beam energy by selecting central events. Ogilvie $\textit{et
al.}$ \cite{ogil91} also studied the multifragment decays of Au
projectiles after collisions with C, Al, and Cu targets at the
bombarding energy of 600 MeV/nucleon using ALADIN forward
spectrometer at GSI, Darmstadt, with the beam accelerated by SIS
synchrotron. They found that with increasing the violence of
collision, the mean multiplicity of IMFs originating from
projectile first increases to a maximum and then decreases again.
\par
As mentioned earlier, Sisan $\textit{et al.}$ \cite{sisan01}
reported that the peak multiplicity of IMFs as well as peak
center-of-mass energy scale with the size of the system. In a
recent communication, Vermani and Puri \cite{puri09} succeeded
partially in explaining the above mentioned behaviour by using the
quantum molecular dynamics (QMD) approach. We here plan to extend
the above study by incorporating various model ingredients such as
equation of state, nucleon-nucleon (nn) cross section, and
Gaussian width. The role of different clusterization algorithms
shall also be explored. We shall attempt to find out whether these
ingredients have sizable effects.
\par
\section{The Formalism}
\subsection{Quantum Molecular dynamics (QMD) model}
\par
We describe the time evolution of a heavy-ion reaction within the
framework of Quantum Molecular Dynamics (QMD) model
\cite{rkpuri,sood09,jsingh00,aich91} which is based on a molecular
dynamics picture. The explicit two- and three-body interactions
lead to the preservation of fluctuations and correlations that are
important for N-body phenomena like multifragmentation. In QMD
model each nucleon is represented by a Gaussian distribution whose
centroid propagates with the classical equations of motion:
\begin{equation}
\frac{d\mathbf{r}_{i}}{dt}=\frac{dH}{d\mathbf{p}_{i}}, \label {e1}
\end{equation}
\begin{equation}
\frac{d\mathbf{p}_{i}}{dt}=-\frac{dH}{d\mathbf{r}_{i}},
\end{equation}
where the Hamiltonian is given by
\begin{equation}
H=\sum_{i} \frac{\mathbf{p}^{2}_{i}}{2m_{i}}+V^{tot},
\end{equation}
with

\begin{equation}
V^{tot} = V^{loc}+  V^{Coul} + V^{Yuk} + V^{MDI}, \label {e10}
\end {equation}
$V^{loc}$ is the Skyrme force whereas $V^{Coul}$, $V^{Yuk}$ and
$V^{MDI}$ define, respectively, the Coulomb, Yukawa and momentum
dependent potentials. Yukawa term separates surface which also
play role in low energy process like fusion and cluster
radioactivity \cite{dutt10,malik}. The momentum-dependent part of
the interaction acts strongly in the cases where the system is
mildly excited \cite{kumarpuri99,kumarpuri98}. In this case, the
MDI is reported to generate a lot more fragments compared to the
static equation of state. For a detailed discussion of the
different equations of state and MDI, the reader is referred to
Refs. \cite{blaich93,kumarpuri99,kumarpuri98}. The relativistic
effect does not play role in low incident energy of present
interest.
\par
The phase space of the nucleons is stored at several time steps.
The QMD model does not give any information about the fragments
observed at the final stage of the reaction. In order to construct
fragments from the present phase-space one needs the
clusterization algorithms. We shall concentrate here on the MST
and MSTB methods only.
\par
\subsection{Different clusterization methods}
\subsubsection{Minimum spanning tree (MST) method}
 The widely used clusterization algorithm is the Minimum
Spanning Tree (MST) method \cite{aich91}. In MST method, two
nucleons are allowed to share the same fragment if their centroids
are closer than a distance $r_{min}$,
\begin{equation}
|\textbf{r}_{\textbf{i}}-\textbf{r}_{\textbf{j}}| \leq r_{min}.
\end{equation}
where $\textbf{r}_{\textbf{i}}$ and $\textbf{r}_{\textbf{j}}$ are
the spatial positions of both nucleons. The value of $r_{min}$ can
vary between 2-4 fm. This method cannot address the question of
time scale. This method gives a big fragment at high density which
splits into several light and medium mass fragments after several
hundred fm/c. This procedure gives same fragment pattern for times
later than 200 fm/c, but cannot be used for earlier times.
\par
\subsubsection{Minimum spanning tree method with binding energy check (MSTB)}
This is an improved version of normal MST method. Firstly, the
simulated phase-space is analyzed with MST method and pre-clusters
are sorted out. Each of the pre-clusters is then subjected to
binding energy check \cite{puri09}:
\begin{equation}
\zeta_{i} =
\frac{1}{N^{f}}\sum_{i=1}^{N^{f}}[\frac{(\textbf{p}_{i}-P_{N^{f}}^{cm})^{2}}{2m_{i}}
+ \frac{1}{2}\sum_{j \neq
i}^{N^{f}}V_{ij}(\textbf{r}_{i},\textbf{r}_{j})]< E_{bind}.
\end{equation}
We take $E_{bind}$ = -4.0 MeV if $N^{f}\geq 3$ and $E_{bind}$ =
0.0 otherwise. Here $N^{f}$ is the number of nucleons in a
fragment and $P_{N^{f}}^{cm}$ is center-of-mass momentum of the
fragment. This is known as Minimum Spanning Tree method with
Binding energy check (MSTB) \cite{puri09}. The fragments formed
with the MSTB method are reliable and stable at early stages of
the reactions.
\par
\section{Results and Discussion}
We have simulated the central reactions of $^{20}$Ne+$^{20}$Ne
(E$_{lab}$ = 10-55 AMeV), $^{40}$Ar+$^{45}$Sc (E$_{lab}$ = 35-125
AMeV), $^{58}$Ni+$^{58}$Ni (E$_{lab}$ = 35-105 AMeV),
$^{86}$Kr+$^{93}$Nb (E$_{lab}$ = 35-95 AMeV),
$^{129}$Xe+$^{118}$Sn (E$_{lab}$= 45-140 AMeV),
$^{86}$Kr+$^{197}$Au (E$_{lab}$= 35-400 AMeV) and
$^{197}$Au+$^{197}$Au (E$_{lab}$ = 70-130 AMeV). The energies are
guided by experiments \cite{tsang93,sisan01,peas94}. For the
present study, we use hard (labeled as Hard), soft (Soft), Hard
with MDI (HMD) and Soft with MDI (SMD) equation of state. We also
use standard energy-dependent Cugnon cross section ($\sigma$$_{nn}
^{free}$) \cite{kumarpuri98} and constant isotropic cross section
of 55 mb strength in addition to two different widths of Gaussian
L = 1.08 and 2.16 fm$^{2}$ (L$^{broad}$). The superscript to the
labels represent cross section. The phase-space is clusterized
using clusterization methods described previously. The reactions
are followed till 200 fm/c but the conclusions do not change when
the reaction is over employing the validity of both algorithms.
%%%%%%%%%%%%%%%%%%%%%%%%%%%%%%%%%%%%%%%%%%%%
\begin{figure}[!t]
\centering \vskip -1.5cm
\includegraphics[angle=0,width=16cm]{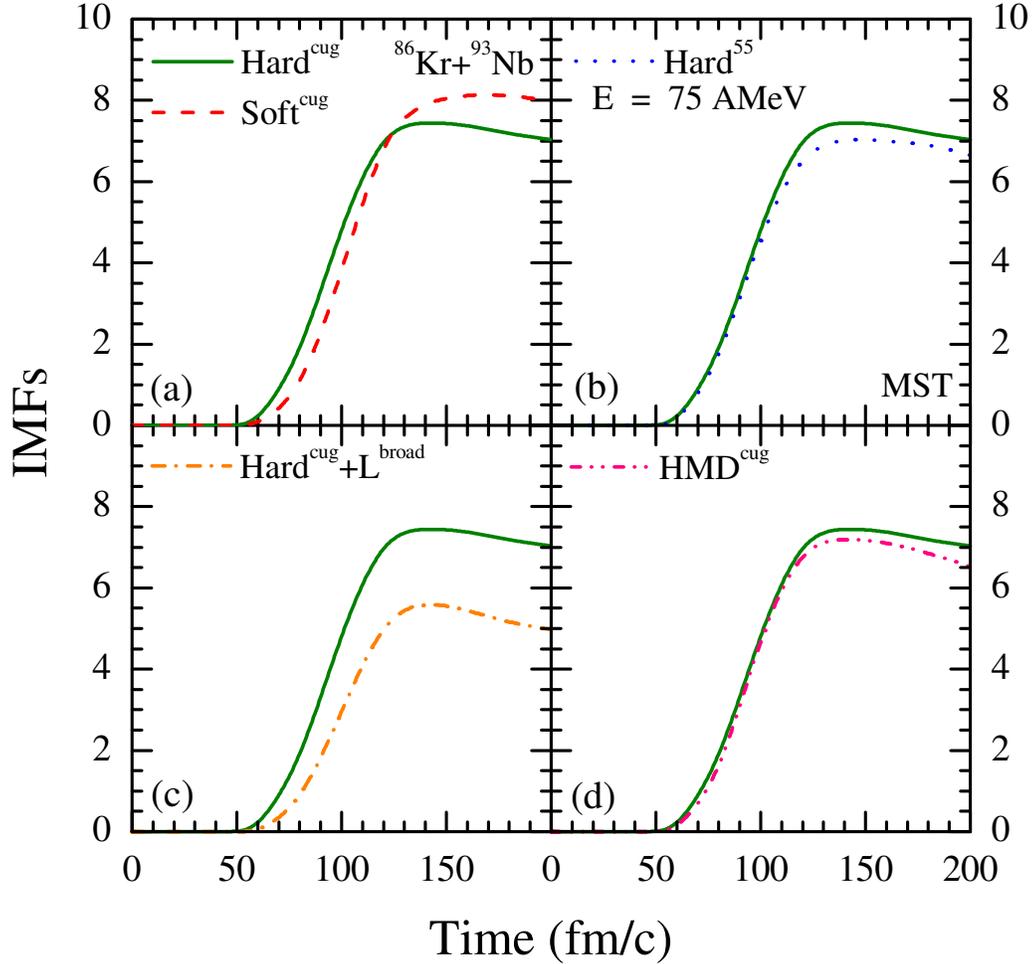}
 \vskip -0cm \caption{ The time evolution of IMFs (5$\leq$A $\leq$44) for the reaction of $^{86}$Kr+$^{93}$Nb at incident energy of 75
AMeV for different model ingredients .}\label{fig1}
\end{figure}
%%%%%%%%%%%%%%%%%%%%%%%%%%%%%%%%%%%%%%%%%%%%%%%%%%
\par
In fig. 1, we display the time evolution of IMFs for the reaction
$^{86}$Kr+$^{93}$Nb at incident energy of 75 AMeV employing MST
method. In fig. 1(a), we display the model calculations using
Hard$^{cug}$ (solid line) and Soft$^{cug}$ (dashed line). From
fig. 1(a), we see that the number of IMFs are larger in case of
Soft as compared to Hard. This is because of the fact that soft
matter will be easily compressed as a result of which density
achieved will be more which in turn will lead to the large number
of IMFs as compared to that in case of Hard. It is worth
mentioning here that the effect could be opposite at higher
energies. Since at higher energies the IMFs may further break into
LCPs and free nucleons. In fig. 1(b), we display the results for
Hard$^{cug}$ and Hard$^{55}$ (dotted line). As evident from the
fig. 1(b), the number of IMFs are nearly same for both type of
cross sections. This may be due to the fact that for the central
collisions, since the excitation energy is already high therefore,
different cross sections have a negligible role to play. In fig.
1(c), we display the results for Hard along with two different
widths of Gaussian i.e. L and L$^{broad}$ (dash-dotted line). We
find that the width of Gaussian has a considerable impact on
fragmentation. As we change the Gaussian width (L) from 4.33 to
8.66 fm$^{2}$, the multiplicity of IMFs is reduced by $\approx$
30$\%$. Interestingly, the kaon yield also get reduced by the same
amount \cite{hart98}. Due to its large interaction range, an
extended wave packet (i.e. L$^{broad}$) will connect a large
number of nucleons in a fragment, as a result it will generate
heavier fragments as compared to one obtains with smaller width.
It is worth mentioning here that the width of Gaussian has a
considerable effect on the collective flow \cite{gautm10,hart98}
as well as pion production also \cite{hart94,hart98}. In fig.
1(d), we display the results using Hard and HMD (dash-dot-dot
line). Again the number of IMFs are nearly same for both EOS. This
is expected since the effect of MDI will be small at these
energies. However, the scenario is completely different at high
energies. Since at high energies, due to the repulsive nature of
MDI, there is large destruction of initial correlations and the
additional momentum dependence further destroys the correlations
reducing further the multiplicity of IMFs. This leads to the
emission of lots of nucleons and LCPs \cite{goyal09}.

\begin{figure}[!t]
\centering \vskip -1.5cm
\includegraphics[angle=0,width=12cm]{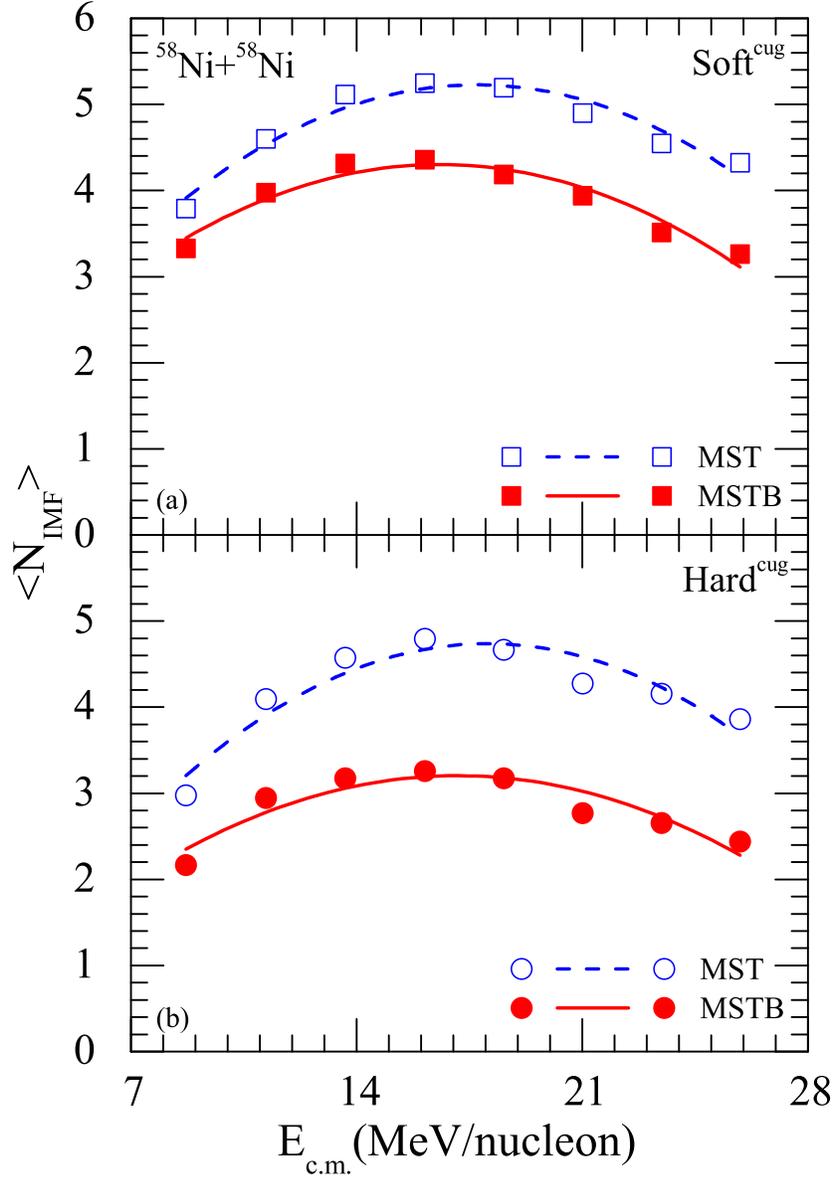}
\vskip 0cm \caption{The mean IMF multiplicity,
$\langle$N$_{IMF}\rangle$, as a function of incident energy in
center-of-mass frame, E$_{c.m.}$, for the reaction of
$^{58}$Ni+$^{58}$Ni. Solid (dashed) curves show the quadratic fits
to the model calculations for MSTB (MST) method to estimate the
peak center-of-mass energy at which the maximal IMF emission
occurs.}\label{fig2}
\end{figure}
\par
In fig. 2, we display the average multiplicity of IMFs,
$\langle$N$_{IMF}\rangle$, as a function of incident energy in the
center-of-mass frame (E$_{c.m.}$) for $^{58}$Ni+$^{58}$Ni reaction
employing MST (open symbols) and MSTB (solid symbols) methods.
Figs. 2(a) and 2(b) are for Soft$^{cug}$ and Hard$^{cug}$,
respectively. Lines represent the quadratic fit to the model
calculations. In both cases, the number of IMFs first increases
with incident energy, attains a maxima and then decreases in
agreement with the previous studies
\cite{tsang93,sisan01,peas94,ogil91,puri09}. Clearly,
$\langle$N$_{IMF}\rangle$ is more for MST method as compared to
MSTB method. Since in case of MSTB method along with spatial
correlations, an additional check for binding energy is also used,
therefore it filters out the loosely bound fragments which will
decay later. Hence, the fragments obtained with MSTB method are
properly bound. A similar trend is obtained for all other
reactions as well as different model ingredients used in the
present study but is less pronounced in lighter systems like
$^{20}$Ne+$^{20}$Ne, $^{40}$Ar+$^{45}$Sc as compared to heavier
systems. However for Gaussian width L$^{broad}$, the
$\langle$N$_{IMF}\rangle$ is nearly zero in this incident energy
range using MSTB method (not shown here). This is due to the fact
that an extended wave packet (i.e. L$^{broad}$) connects a large
number of nucleons in a fragment, as a result it generates heavier
fragments and the additional binding energy check further excludes
the unbound fragments.

\begin{figure}[!t]
\centering \vskip-2cm
\includegraphics[angle=0,width=16cm]{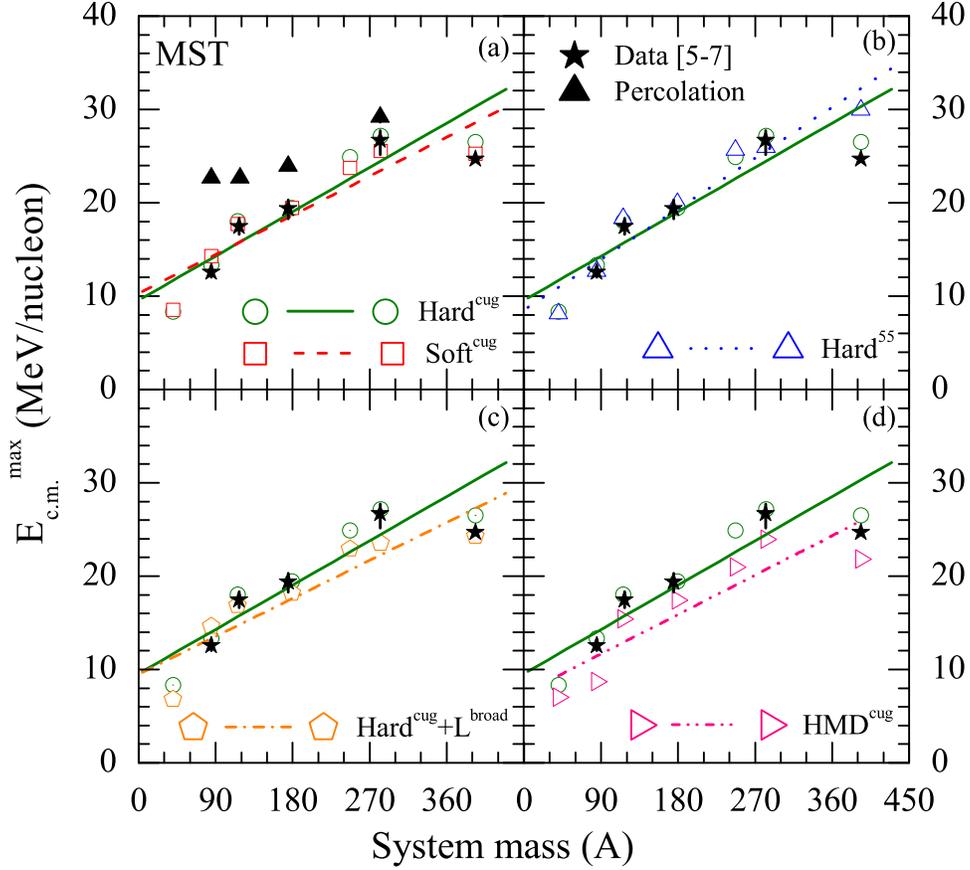}
\vskip -6cm \caption{The E$_{c.m.} ^{max}$ as a function of
composite mass of the system (A). The different lines represent
the linear fits. Comparison of model calculations is made with
experimental data \cite{tsang93,sisan01,peas94} (solid stars). The
percolation calculations \cite{sisan01} (solid triangles) are also
shown in figure.}\label{fig3}
\end{figure}
\par
In fig. 3, we display the peak center-of-mass energy E$_{c.m.}
^{max}$ as a function of combined mass of the system employing MST
method. Lines represent linear fitting proportional to mA. We find
that the mass dependence of E$_{c.m.} ^{max}$ is insensitive to
different EOS (fig. 3a), nn cross section (fig. 3b) as well as the
width of Gaussian also (fig. 3c). It is slightly sensitive to MDI
because for heavy systems E$_{c.m.} ^{max}$ is more as a result of
which the effect of MDI becomes non-negligible. In fig. 3, the
model calculations are also compared with experimental data
\cite{tsang93,sisan01,peas94}. It is clear from the fig. 3 that
model calculations for E$_{c.m.} ^{max}$ agree with experimental
data \cite{tsang93,sisan01,peas94}. This behavior is consistent
for all the different choices of model ingredients.

\begin{figure}[!t]
\centering \vskip -2cm
\includegraphics[angle=0,width=16cm]{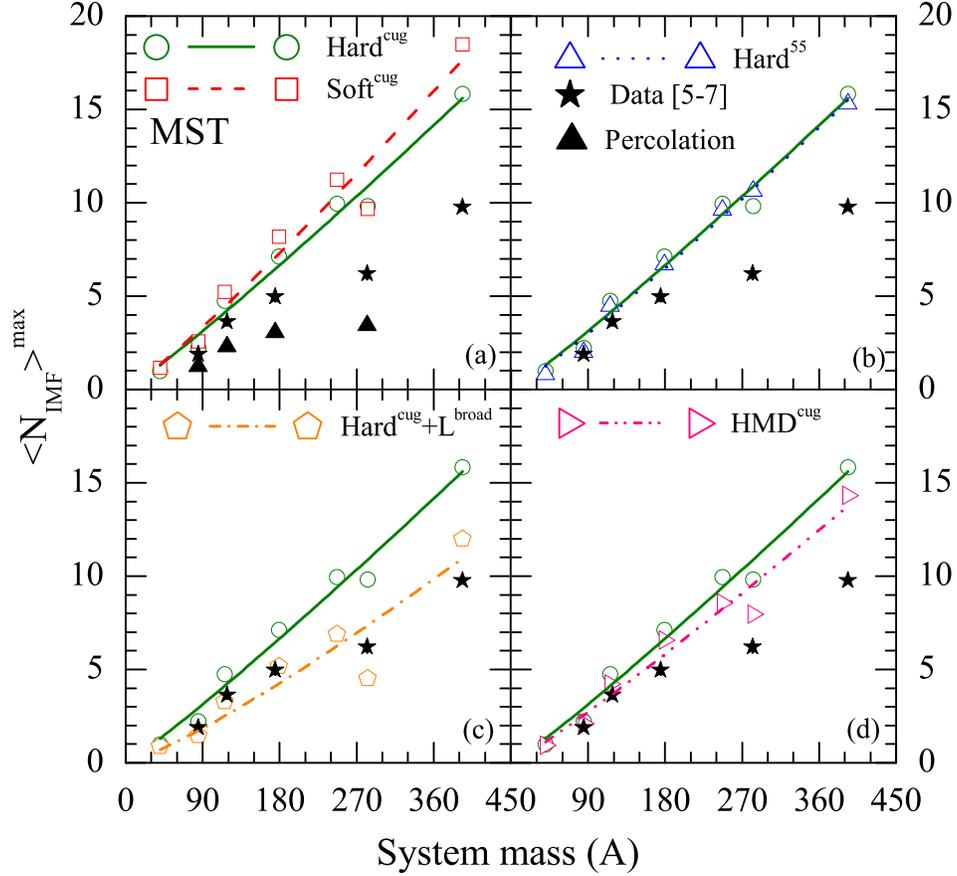}
\vskip -6cm \caption{$\langle$N$_{IMF}\rangle^{max}$  as a
function of composite mass of the system (A). The different lines
represent the power law fits ($\propto$ A$^{\tau}$). Comparison of
model calculations is made with experimental data
\cite{tsang93,sisan01,peas94} (solid stars). The percolation
calculations \cite{sisan01} (solid triangles) are also shown in
figure.}\label{fig4}
\end{figure}

\par
In fig. 4, we display the peak multiplicity of IMFs
$\langle$N$_{IMF}\rangle^{max}$ as a function of combined mass of
the system employing MST method. The lines represent power law
fitting proportional to A$^{\tau}$. The multiplicity of IMFs, in
case of $^{20}$Ne+$^{20}$Ne and $^{40}$Ar+$^{45}$Sc, is obtained
by excluding the largest and second largest fragment,
respectively, to get the accurate information about the system
size dependence. $\langle$N$_{IMF}\rangle^{max}$ are obtained by
making a quadratic fit to the model calculations for
$\langle$N$_{IMF}\rangle$ as a function of (E$_{c.m.}$). We find
that the peak multiplicity is insensitive to cross section (fig.
4b) and MDI (fig. 4d) (for explanation see discussion of fig. 1).
It is slightly sensitive to EOS (fig. 4a but highly sensitive to
the Gaussian width (fig. 4c). On increasing the width of Gaussian,
$\langle$N$_{IMF}\rangle^{max}$ reduces to a large extent. As
discussed earlier, an extended wave packet (i.e. L$^{broad}$) will
connect a large number of nucleons in a fragment, as a result it
generates heavier fragments as compared to one obtains with
smaller width. From fig. 3, we see that E$_{c.m.} ^{max}$ shows
linear dependence ($\propto$ mA) whereas
$\langle$N$_{IMF}\rangle^{max}$ (fig. 4) follows power law
behaviour ($\propto$ A$^{\tau}$) with $\tau$ nearly equal to
unity. In fig. 4, the model calculations are also compared with
experimental data \cite{tsang93,sisan01,peas94}. It is clear from
the fig. 4 that, as the system mass increases difference between
model calculations and experimental results goes on increasing.
This behavior is consistent for all the different choices of model
ingredients. This may be due to the fact that the fragments
obtained with MST method are not reliable because this method
makes sense only when matter is diluted and well separated. This
is true only in case of high beam energy and in central
collisions. Therefore, we have to look for other methods of
clusterization. As mentioned earlier, the fragments obtained with
MSTB method are properly bound and reliable. So, as a next step,
we check system size dependence of E$_{c.m.} ^{max}$ and
$\langle$N$_{IMF}\rangle^{max}$ by using MSTB method for
clusterization.
\par

\par
\begin{figure}[!t]
\centering \vskip -1.5cm
\includegraphics[angle=0,width=16cm]{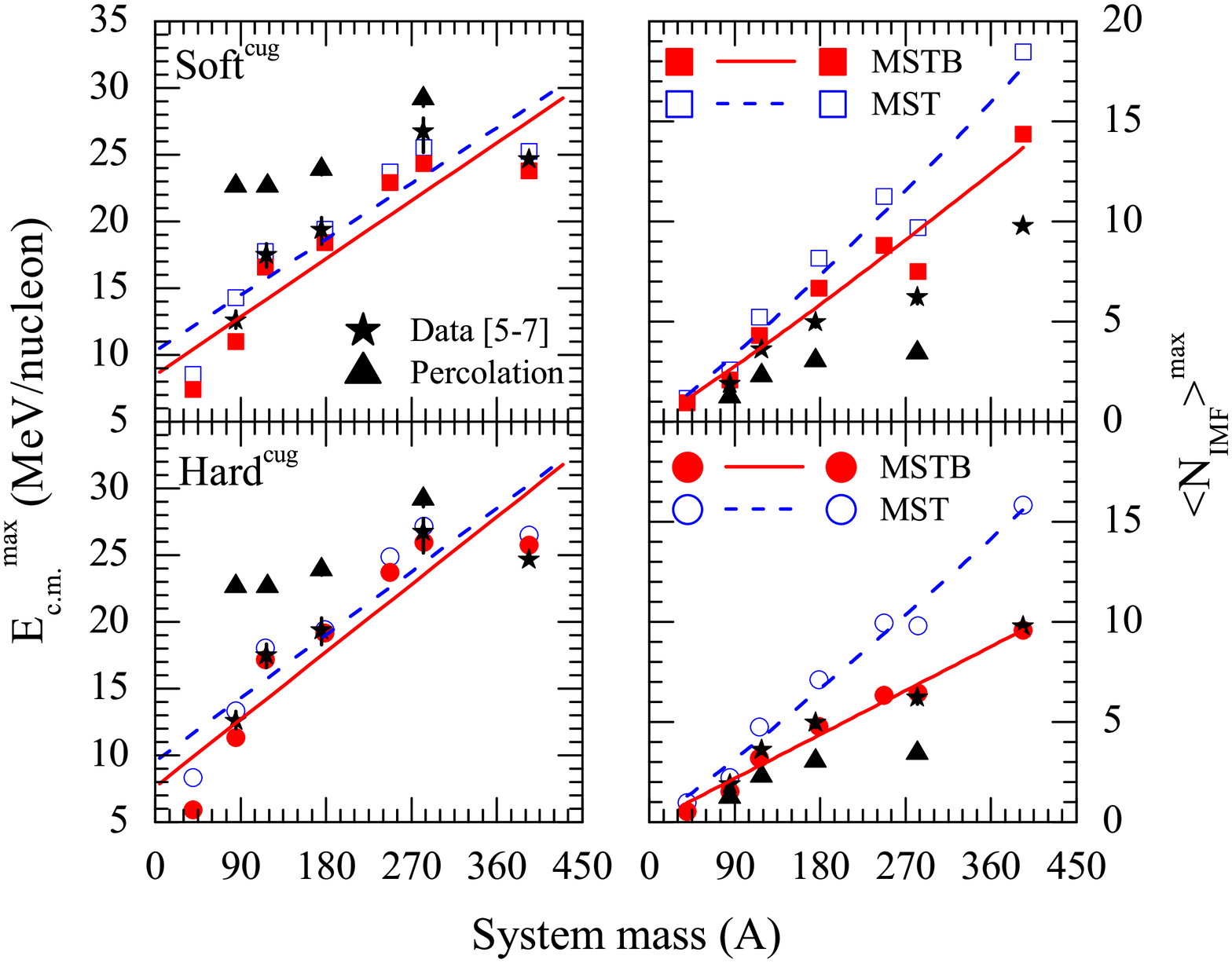}
\vskip -6cm \caption{The E$_{c.m.} ^{max}$ (left panels) and
$\langle$N$_{IMF}\rangle^{max}$ (right panels) as a function of
composite mass of the system (A) using Soft$^{cug}$ (upper panels)
and Hard$^{cug}$ (lower panels) employing MSTB and MST methods.
The different lines in left (right) panels represent the linear
fits (power law fits). Comparison of model calculations is made
with experimental data \cite{tsang93,sisan01,peas94} (solid
stars).}\label{fig5}
\end{figure}
In fig. 5, we display the E$_{c.m.} ^{max}$ (left panels) and
$\langle$N$_{IMF}\rangle^{max}$ (right panels) for Soft$^{cug}$
(upper panels) and Hard$^{cug}$ (bottom panels) as a function of
combined mass of the system. Solid (open) symbols represent MSTB
(MST) method. From left panels we find that E$_{c.m.} ^{max}$
remains insensitive to the choice of clusterization method. The
same is true for $\langle$N$_{IMF}\rangle^{max}$ (right panels)
but in low mass region. As the system mass increases, the
$\langle$N$_{IMF}\rangle^{max}$ becomes more and more sensitive to
the method of clusterizaton. The MSTB method excludes the loosely
bound fragments thus reducing the peak IMF multiplicity. The
effect is uniform for both the EOS as well as for different cross
section (not shown here).
\par
\begin{figure}[!t] \centering
 \vskip -1.5cm
\includegraphics[angle=0,width=12cm]{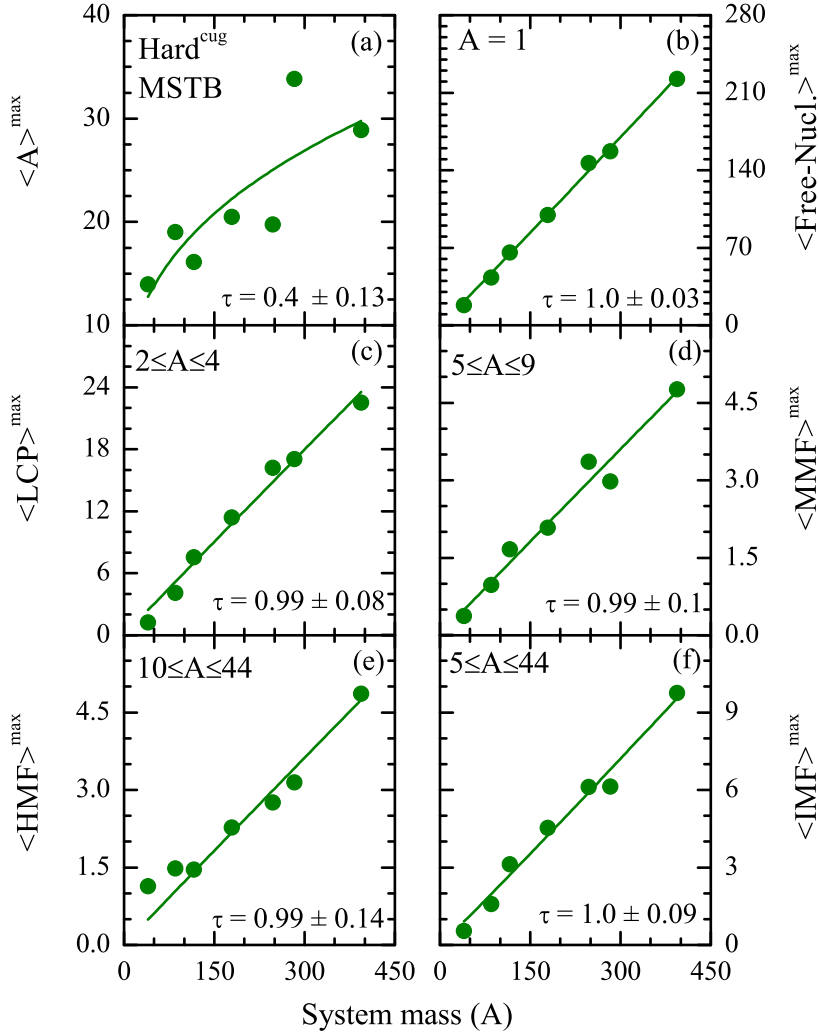}
 \vskip -1.5cm \caption{ The largest fragment and multiplicities of free-nucleons, LCPs, MMFs, HMFs, and IMFs as a
 function of composite mass of the colliding nuclei (A) for different reactions at their respective E$_{c.m.}^{max}$ (solid circles).
 Lines represent the power law fits ($\propto$ A$^{\tau}$).}\label{fig6}
\end{figure}
\par
In fig. 6, we display peak multiplicity (obtained by employing
MSTB method) as a function of composite mass of the system for
various fragments consisting of the largest fragment (A$^{max}$)
(fig. 6a), free-nucleons (1$\leq$A $\leq$1) (fig. 6b), light
charged particles (LCPs) (2$\leq$A $\leq$4) (fig. 6c), medium mass
fragments (MMFs) (5$\leq$A $\leq$9) (fig. 6d), heavy mass
fragments (HMFs) (10$\leq$A $\leq$44) (fig. 6e) and intermediate
mass fragments (IMFs) (5$\leq$A $\leq$44) (fig. 6f) for
Hard$^{cug}$. Lines represent the power law fitting proportional
to A$^{\tau}$. Interestingly, the peak multiplicities of different
fragments follow a power law ($\propto$ A$^{\tau}$). Power law
factor $\tau$ is almost unity in all cases except A$^{max}$ for
which there is no clear system size dependence. The system size
dependence of various fragments has also been predicted by Dhawan
and Puri \cite{dhawan06}. Their calculations at the energy of
vanishing flow (i.e., the energy at which the transverse flow
vanishes) clearly suggested the existence of a power law system
mass dependence for various fragment multiplicities.

\par
\section{Summary}
We have simulated the central reactions of nearly symmetric, and
asymmetric systems over the entire periodic table at different
incident energies for the different equations of state (EOS), nn
cross sections and different widths of Gaussians. We have observed
that the multiplicity of intermediate mass fragments (IMFs)
($3\leq$Z$\leq20$) shows a rise and fall with increase in beam
energy in the center-of-mass frame as already predicted
experimentally/theoretically. We have also studied the system size
dependence of peak center-of-mass energy E$_{c.m.} ^{max}$ and
peak IMF multiplicity $\langle$N$_{IMF}\rangle^{max}$. It has been
observed that E$_{c.m.}^{max}$ increases linearly with system mass
whereas a power law ($\propto$ A$^{\tau}$) dependence has been
observed for $\langle$N$_{IMF}\rangle^{max}$ with $\tau\sim$1.0.
We have compared system size dependence of E$_{c.m.} ^{max}$ and
$\langle$N$_{IMF}\rangle^{max}$
 for MST and MSTB methods and found that MSTB method reduces the $\langle$N$_{IMF}\rangle^{max}$ especially in heavy systems because in MSTB
 method due to binding energy check loosely bound fragments get excluded.
 The power law dependence is also observed for fragments of
different sizes at the energy for which the production of IMFs is
maximum and power law parameter $\tau$ is found to be close to
unity in all cases except A$^{max}$.
\par
\section{Acknowledgements}
 This work has been supported by a grant from Department of
Science and Technology (DST), Government of India, India.

\end{document}